\documentclass[aps,pra,showpacs]{revtex4}
\usepackage{amsmath}
\usepackage{amssymb}
\newcommand\Ho{{\hat H}}
\newcommand\half{{\frac{1}{2}}}
\newcommand\Pb{\mathbf{P}}

\newcommand\kb{\mathbf{k}}
\newcommand\Qb{\mathbf{Q}}
\newcommand\Eo{{\hat E}}
\newcommand\ro{{\hat\rho}}
\newcommand\kbo{{\mathbf{\hat k}}}
\newcommand\Pbo{{\mathbf{\hat P}}}
\newcommand\Xbo{{\mathbf{\hat X}}}
\newcommand\Vo{{\hat V}}
\newcommand\Lc{\mathcal{L}}
\def\ket#1{\vert#1\rangle}
\def\bra#1{\langle#1\vert}
\newcommand\To{{\hat T}}

\begin{document}
\title{Quantum linear Boltzmann equation with finite intercollision time}    
\author{Lajos Di\'osi}
\email{diosi@rmki.kfki.hu}
\homepage{www.rmki.kfki.hu/~diosi} 
\affiliation{
Research Institute for Particle and Nuclear Physics\\
H-1525 Budapest 114, POB 49, Hungary}
\date{\today}

\begin{abstract}
Inconsistencies are pointed out in the usual quantum versions of the classical linear Boltzmann 
equation constructed for a quantized test particle in a gas. These are related to the incorrect formal
treatment of momentum decoherence. We prove that ideal collisions would result in complete momentum
decoherence, the persistence of coherence is only due to the finite intercollision time.
A corresponding novel quantum linear Boltzmann equation is proposed.
\end{abstract}

\maketitle
\emph{Introduction.}
Collisional theory of motion of a test particle through a gas had been the oldest classical tool  
confirming atomic kinetics - e.g. in Brownian motion \cite{Ein05} - or using atomic kinetics - 
e.g. Millikan's experiment \cite{Mil23}.      
If the gas is dilute and hot, the interaction of the particle with the gas reduces to the repeated 
independent binary collisions between the particle and the individual molecules. 
The dynamics of the test particle is fully captured by the differential cross section of the scattering: 
the linear Boltzmann equation (LBE) describes the particle's motion.  
For a quantum particle, the scattering matrix $\hat 1 +i\To$ is expected to 
capture the dynamics which we describe by the quantum linear Boltzmann equation (QLBE):
\begin{equation}\label{QLBE}
\frac{d\ro}{dt}=-i[\Ho,\ro]+\Lc\ro~,
\end{equation} 
where $\ro$ is the density matrix, $\Ho$ is the free Hamiltonian (including an energy shift due to the gas) 
and $\Lc$ is the Lindblad \cite{LinKosetal76} superoperator of the non-Hamiltonian evolution.
The microscopic derivation of $\Lc$ does not differ from the classical derivation of the LBE as long 
as we assume a density matrix $\bra{\Pb}\ro\ket{\Pb^\prime}=\rho(\Pb)\delta(\Pb-\Pb^\prime)$ without superpositions in momentum $\Pb$.
The evolution of the diagonal density matrix is governed by the classical LBE applied to the distribution $\rho(\Pb)$.
This is why the classical kinetic theory has so far served as a perfect effective theory for a quantum particle in innumerable
applications.   
A stealthy request for a systematic theory emerged in the field of foundations where quantum Brownian motion became
the paradigm of how the environment (gas) `measures' the wave function (of the particle) \cite{JooZeh85,GalFle90,Dio94,DodHal03}. 
For experiments, a proper quantum Brownian particle was not available before, now it is, cf. e.g. \cite{Horetal03}.

When we are interested in local quantum dynamics of the particle we need the off-diagonal elements as well, and then the
microscopic derivation of $\Lc$ becomes problematic. The classical derivation resists to a direct quantum extension. 
The first microscopic derivation was obtained fifteen years ago \cite{Dio95} by the present author. Subsequent 
works \cite{Vac01,Hor06} by Vacchini and Hornberger - through transients \cite{HorSip03} and rectifications \cite{Adl06} -
led them to a refined version \cite{HorVac08}. Their up-to-date review \cite{VacHor09} further elucidates the 
subject and can serve as an exhaustive source of references for the reader. 
Microscopic derivations assume typically (though not necessarily) the Maxwell-Boltzmann distribution for the
molecular momentum $\kb$:
\begin{equation}\label{rhog}
\rho_g(\kb)=\left(\frac{\beta}{2\pi m}\right)^{3/2}\exp\left(-\frac{\beta\kb^2}{2m}\right)~,
\end{equation}
where $m$ is the molecular mass, $\beta$ is the inverse temperature. To overcome the mentioned difficulties with the off-diagonal
elements of $\ro$, Ref.~\cite{Dio95} suggested a formal resolution which became an irreducible part of all subsequent
derivations \cite{HorSip03,Adl06,Vac01,Hor06,HorVac08,VacHor09}. In all variants of QLBE, including a most recent one \cite{KamCre09}, $\Lc$ is not 
linear functional of $\rho_g(\kb)$ as it should be, rather $\Lc$ is a quadratic functional of the \emph{square-root} of $\rho_g(\kb)$. 
In other words: $\Lc$ is the linear functional of 
\begin{equation}\label{sqrt}
\sqrt{\rho_g(\kb)\rho_g(\kb^\prime)}~.  
\end{equation}
Therefore, no matter which QLBE is considered, we face the following issue of consistency. Suppose the gas has two
independent `components' consisting of the same molecules at same densities while at two different temperatures. 
Then their contribution to the QLBE must be additive. But it is not, since $\Lc$ in QLBE (\ref{QLBE}) is linear in 
the expression (\ref{sqrt}), not in $\rho_g(\kb)$ itself. 

Before we construct a novel QLBE which is linear in $\rho_g(\kb)$, we point out a surprising collisional decoherence effect 
apparently overlooked for longtime and discovered, independently from the present work, by Kamleitner and Cresser, too \cite{KamCre09}. 
It may radically alter our understanding what QLBE should be.

\emph{Single collision decoheres momentum.}
Let us consider the kinematics of a single collision between the particle and a molecule. 
As a Gedanken-experiment, suppose we prepared the molecule with momentum $\kb$ before the collision and we detected it
with momentum $\kb-\Qb$ after the collision, i.e., we know the momentum transfer $\Qb$.
Note that the interactions of these preparation and detection devices with the molecule are irrelevant for the 
reduced dynamics of the particle, yet such preparation and detection devices are relevant for our arguments. 
We get dramatic conclusions for the reduced state $\ro$ of the particle since,
for elementary kinematic reasons, the observation of $\Qb$ is equivalent with the observation of one component of the particle's
momentum $\Pb$. Let $\Pb$ stand for the post-collision momentum, then from energy conservation
\begin{equation}\label{Econs}
\frac{\kb^2}{2m}+\frac{(\Pb-\Qb)^2}{2M}=\frac{(\kb-\Qb)^2}{2m}+\frac{\Pb^2}{2M}~
\end{equation}
we have the following relationship:
\begin{equation}\label{Ppar}
\Pb_{\Vert}=\frac{M}{m}\kb_\Vert - \left(\frac{M}{m}-1\right)\frac{\Qb}{2}~,
\end{equation}
where the subscripts $\Vert$ refer to the components parallel to $\Qb$. It is obvious that after our single collision 
the particle's density matrix $\ro$, whatever it was before the collision, becomes perfect diagonal
in $\Pb_\Vert$. Gradually after many collisions the state $\ro$ becomes a mixture of plane waves,
no off-diagonal mechanism will be left at all.
This result contradicts to what we have so far learned as QLBE. The art of deriving a plausible QLBE has always been about the nontrivial
treatment of the off-diagonal elements. The above decoherence mechanism, however, shows that standard scattering theory and 
independent collisions do not yield any QLBE for the off-diagonals. The apparent off-diagonal dynamics may well have been 
artifacts of the square-root fenomenology (\ref{sqrt}).       
     
In realty, the state of the particle is not necessarily the delocalized one advocated by our decoherence argument.
In fact, our argument is relevant for the case of very dilute gas or for very small collision cross sections. 
There the density matrix has no off-diagonal elements at all. Such perfect momentum decoherence assumes perfect energy conservation 
of the collision process. In quantum scattering it requires infinite time and this is the loophole where we concentrate. 
If the intercollision time is moderate long it still leaves ample room for energy balance uncertainty and momentum coherence. 

\emph{QLBE with finite intercollision time.}
We outline the derivation of QLBE (\ref{QLBE}) where $\Lc$ remains linear functional of $\rho_g(\kb)$ and momentum decoherence is
described correctly in function of the intercollision time $\tau$.
The full quantum derivation introduces the box-normalized single-molecule density matrix:  
\begin{equation}\label{rog}
\ro_g=\frac{(2\pi)^3}{V}\rho_g(\kbo)~.
\end{equation}
We follow Ref.~\cite{Dio95}.
Starting from the pre-collision initial state $\ro\otimes\ro_g$, we calculate the collisional change of the particle's reduced state.
The part which is responsible for in-scattering is given by the following expression:
\begin{equation}\label{roin}
\Delta\ro\vert_{in}=Vn_g\mathrm{tr}_g({\To}\ro\otimes\ro_g{\To}^\dagger)~, 
\end{equation}
where the single-collision contribution has been enhanced by the total number $Vn_g$ of molecules at density $n_g$.  
The standard form of the transition operator for the (spin-independent) scattering with initial laboratory momenta
$\kb$ and $\Pb$, resp., reads: 
\begin{equation}\label{To}
\To=\frac{1}{2\pi m^\ast}\int d\kb d\Qb \mathrm{e}^{i\Qb\Xbo} 
        f(\kbo_f^\ast,\kbo_i^\ast)
        \delta(\Eo_{fi}^\ast)
        \ket{\kb-\Qb}\bra{\kb}~.
\end{equation}
The function $f$ is the scattering amplitude. The quantities marked by $\ast$ are the initial and final 
(pre- and post-collision) c.o.m. momenta, while $\Eo_{fi}^\ast$ is the c.o.m. energy balance, resp.:
\begin{eqnarray}\label{com}
\kbo_i^\ast&=&\frac{M}{M^\ast}\kb-\frac{m}{M^\ast}\Pbo~,\nonumber\\
\kbo_f^\ast&=&\kbo_i^\ast-\Qb\nonumber~,\\
\Eo_{fi}^\ast&=&\frac{1}{2m^\ast}(\kbo_f^\ast)^2-\frac{1}{2m^\ast}(\kbo_f^\ast)^2~.
\end{eqnarray}
Here $M^\ast=M+m$ and $m^\ast=Mm/M^\ast$.
We substitute the Eqs.~(\ref{rog}) and (\ref{To}) into (\ref{roin}), yielding:
\begin{equation}\label{roin1}
\Delta\ro\vert_{in}
=\frac{2\pi n_g}{m^{\ast2}}\int\int d\kb\rho_g(\kb) d\Qb 
\mathrm{e}^{i\Qb\Xbo}f(\kbo_f^\ast,\kbo_i^\ast)\delta(\Eo_{fi}^\ast)\ro 
{\bar f}(\kbo_f^\ast,\kbo_i^\ast)\delta(\Eo_{fi}{^\ast\prime})\mathrm{e}^{-i\Qb\Xbo}~. 
\end{equation}
And now, we alter (and simplify) the old derivation \cite{Dio95}. 
In standard quantum scattering theory, both the preparation of pre-collision and the emergence of post-collision states take infinite 
long time and the transition operator (\ref{To}) contains the $\delta$-function for perfect energy conservation. In our case, 
however, the intercollision time $\tau$ is finite:
\begin{equation}\label{tau}
\tau=\frac{\sqrt{\pi\beta m}}{\sigma n_g}~,
\end{equation}
$\sigma$ is the total scattering cross section. 
We take this feature into the account if in $\To$ we use the
``smoothened'' $\delta$-function $\delta_\tau(E)$ instead of $\delta(E)$:
\begin{equation}\label{deltatau}
\delta_\tau(E)=\frac{\sin(\tau E/2)}{\pi E}~,
\end{equation}
and we extend the transition amplitude $f(\kb_f^\ast,\kb_i^\ast)$ off-shell. 
We shall approximate $d\ro/dt\vert_{in}$ by $\Delta\ro\vert_{in}/\tau$, yielding:
\begin{equation}\label{droindt}
\frac{d\ro}{dt}{~\atop \vert_{in}}
=\frac{2\pi n_g}{\tau m^{\ast2}}\int\int d\kb \rho_g(\kb) d\Qb 
\mathrm{e}^{i\Qb\Xbo} f(\kbo_f^\ast,\kbo_i^\ast)\delta_\tau(\Eo_{fi}^\ast)\ro
                          {\bar f}(\kbo_f^\ast,\kbo_i^\ast)\delta_\tau(\Eo_{fi}^\ast)\mathrm{e}^{-i\Qb\Xbo}~. 
\end{equation}
This form explicitly indicates the following Lindblad structure of our novel QLBE:
\begin{equation}\label{QLBEnov}
\frac{d\ro}{dt}=-i[\Ho,\ro]+\frac{2\pi n_g}{\tau m^{\ast2}}\int\int d\kb \rho_g(\kb) d\Qb
\left(\Vo(\kb,\Qb)\ro\Vo^\dagger(\kb,\Qb)-\half\{\Vo^\dagger(\kb,\Qb)\Vo(\kb,\Qb)~,\ro\}\right)~,
\end{equation}
where the Lindblad operators read:
\begin{equation}\label{VkQ}
\Vo(\kb,\Qb)=\mathrm{e}^{i\Qb\Xbo} f(\kbo_f^\ast,\kbo_i^\ast)\delta_\tau(\Eo_{fi}^\ast)~.
\end{equation}

We expect the following features.
In very dilute gas, we can take $\tau\rightarrow\infty$ and $\delta_\tau\rightarrow\delta$, the off-diagonal elements of $\ro$
become completely suppressed and our QLBE recovers the standard LBE for the diagonal elements of the density matrix. 
If the gas is less dilute, still the LBE remains valid for the diagonal part of $\ro$ while the off-diagonal part reappears and
then $\tau$ becomes the relevant quantity for it. We are going to illustrate these features. 

\emph{Diffusion limit.}
A common test of QLBE is Dekker's equation \cite{Dek81} containing the fenomenological constants $\eta$, $D_{pp}$, $D_{xx}$ of friction, 
momentum and position diffusion (i.e.: position and momentum decoherence) resp.:   
\begin{equation}\label{Dek}
\frac{d\ro}{dt}=-i[\Ho,\ro]-D_{pp}[\Xbo,[\Xbo,\ro]]-D_{xx}[\Pbo,[\Pbo,\ro]]-i\frac{\eta}{2M}[\Xbo,\{\Pbo,\ro\}]~.
\end{equation}
For a heavy Brownian particle $(M\gg m)$, in the diffusion limit (cf., e.g. \cite{Dio95,VacHor09}) this equation
is recovered by all QLBEs. They yield the same (classical) momentum diffusion constant $D_{pp}$ and friction constant $\eta$:
\begin{equation}\label{etaDpp}
\beta\eta=D_{pp}=\frac{1}{6}
\frac{n_g}{m}\int\int d\Qb Q d\kb_\bot\rho_g(\kb_\bot+\Qb/2)
\vert f(\kb_\bot-\Qb/2,\kb_\bot+\Qb/2)\vert^2~.
\end{equation}
Different QLBEs do not agree upon the value of the position diffusion (momentum decoherence) constant $D_{xx}$.
Our QLBE (\ref{QLBEnov}) provides a $\tau$-dependent $D_{xx}$. We are going to derive it.

Since the Dekker Eq.~(\ref{Dek}) corresponds to the 2nd order Taylor expansion of the QLBE (\ref{QLBEnov}) in the operators $\Xbo,\Pbo$, 
we shall compare the respective coefficients of the term $\Pbo\ro\Pbo$ in the Dekker equation and in the QLBE. 
In the latter, the leading term comes from the expansion of the $\delta_\tau(\Eo_{fi}^\ast)$ factors around $\Pbo=0$. 
To calculate their argument for $m/M\rightarrow\infty$, recall Eq.~(\ref{com}):
\begin{equation}\label{Efi}
\Eo_{fi}^\ast=\frac{\Qb}{m}\left(\kb-\frac{\Qb}{2}\right)-\frac{1}{M}\Qb\Pbo\equiv E_{fi}-\frac{1}{M}\Qb\Pbo~.
\end{equation}
Therefore we get:
\begin{equation}
\frac{1}{2}D_{xx}\Pbo\ro\Pbo=\frac{1}{M^2}\frac{2\pi n_g}{\tau m^2}\int\int d\Qb d\kb\rho_g(\kb)
\vert f(\kb-\Qb,\kb)\vert^2 \left[\delta_\tau^\prime(E_{fi})\right]^2 \Qb\Pbo\ro\Pbo\Qb~,
\end{equation}
where we set $\Pbo=0$ in the arguments of $f$ assuming a slowly varying $f$.
Now we change the integration variable according to $dk_\Vert=(m/Q)dE_{fi}$ and perform the integral over $E_{fi}$.
We can ignore the variation of $\rho(\kb)$ and $f(\kb-\Qb,\kb)$ with the variation of $k_\Vert$ which can only take values
from a narrow vicinity of $Q/2$ because of the narrow support of the $\delta_\tau^\prime$. 
Hence, we substitute $\kb_\Vert\approx\Qb/2$ in $\rho_g$ and $f$, then we perform the integral:
\begin{equation}
\int\left[\delta_\tau^\prime(E_{fi})\right]^2 d k_\Vert = 
\frac{m}{Q}\int\left[\delta_\tau^\prime(E_{fi})\right]^2 d E_{fi} =
\frac{m}{Q}\frac{\tau^3}{24\pi}~.
\end{equation}
The resulting equation is the following:
\begin{equation}
\frac{1}{2}D_{xx}\Pbo\ro\Pbo=\frac{\tau^3}{24\pi}\frac{1}{M^2}
\frac{2\pi n_g}{\tau m}\int\int d\Qb d\kb_\bot\rho_g(\kb_\bot+\Qb/2)
\vert f(\kb_\bot-\Qb/2,\kb_\bot+\Qb/2)\vert^2 \frac{1}{Q}\Qb\Pbo\ro\Pbo\Qb~,
\end{equation}
where rotational symmetry allows for substituting $\Qb\Pbo\ro\Pbo\Qb$ by $(Q^2 /3)\Pbo\ro\Pbo$. We get finally that
the quantum position diffusion (momentum decoherence) is proportional to the classical momentum diffusion (\ref{etaDpp}): 
\begin{equation}
D_{xx}=\frac{1}{3}\left(\frac{\tau}{M}\right)^2 D_{pp}~.
\end{equation}
This result is in agreement with the conclusion of our Gedanken-experiment: 
the longer the intercollision time, the stronger the momentum decoherence is.
According to a universal constraint \cite{LinKosetal76,Dek81} of complete positivity of the Dekker Eq.~(\ref{Dek}), the inequality 
$D_{xx}\geq(\beta/4M)^2 D_{pp}$ must be satisfied
which puts a lower bound on position diffusion at fixed friction $\eta$. The QLBE introduced in Ref.~\cite{Hor06} 
predicts this minimum position diffusion. All other previous QLBEs 
\cite{Dio95,HorSip03,Adl06,Vac01,Hor06,HorVac08,VacHor09,KamCre09}
predict similar position diffusion suggesting that it is irrelevant at
high temperatures. Our result is different: position diffusion (momentum decoherence) goes with $\tau^2$, not with $\beta^2$.
The above complete-positivity condition gives the following constraint for $\tau$: 
\begin{equation}
\tau k_B T \geq \sqrt{3}\hbar/4~,
\end{equation}
where $T$ is the temperature; we restored the standard physical dimensions. Substituting Eq.~(\ref{tau}), we get
\begin{equation}
\mathrm{const.}\times\sqrt{m k_B T}/\hbar \geq \sigma n_g~,
\end{equation}
suggesting that our QLBE works until two molecules that hit the particle one after the other get closer than their thermal de Broglie
wave length. This may happen when the density $n_g$ is too high, or the temperature is too low, or the cross section (e.g.: the size of
the particle) is too big. But otherwise, there is an ample region for our particle to develop a localized wave function subject to
our QLBE.     

\emph{Summary.}
Quantum dynamics of a particle through a dilute gas is definitely tractable by the collision model. We pointed out - in agreement with 
Ref.~\cite{KamCre09} - that the coherence between different momentum eigenstates is heavily suppressed and the very persistence 
of any localized coherent dynamics is only due to the uncompleted quantum collisions, i.e., to the finite intercollision time $\tau$.
This brings a dramatic change in our understanding the quantum behavior of the test particle. Previously, momentum
decoherence (i.e.: position diffusion) was considered a weak quantum effect irrelevant at high temperatures. Now we must claim that
such weak momentum decoherence predicted by the previous QLBEs was an artifact. Momentum decoherence is 
a dominating effect. In very dilute gas, at very small scattering rates, the true dynamics is just the classical diffusion (cf. LBE) 
through the momentum eigenstates. This means, in particular, that a gradual momentum decoherence under controlled conditions
may become testable e.g. in fullerene \cite{Horetal03} or in nanoobject interference experiments \cite{Jacetal99Beretal06}. 

There is no guarantee that momentum decoherence has a universal Markovian 
description. The proposed novel QLBE captures the effect of finite intercollision time $\tau$ within the concept 
of independent collisions while the mechanism is beyond it: the second collision literally interferes with the first one! Further
investigations are requested to see whether our QLBE with the finite collision time represents progressive fenomenology compared
to the earlier QLBEs. 

This work was supported by the Hungarian OTKA Grants No. 49384, 75129.

\vfill

\begin{thebibliography}{99}
\bibitem{Ein05} A. Einstein, Ann. Phys. (Leipzig) {\bf 322}, 549.
\bibitem{Mil23} R. A. Millikan, Phys. Rev. {\bf 22}, 1 (1923).
\bibitem{LinKosetal76} G. Lindblad, Comm. Math. Phys. {\bf 48} 119 (1976);
                 V. Gorini, A. Kossakowski, E. C. G. Sudarshan, J. Math. Phys. {\bf 17} 821 (1976).
\bibitem{JooZeh85} E. Joos and H. D. Zeh, Z. Phys. {\bf B59}, 223 (1985) 
\bibitem{GalFle90} M. R. Gallis and G. N. Fleming, Phys. Rev. {\bf A42}, 38 (1990). 
\bibitem{Dio94} L.Di\'osi: {\it Quantum particle as seen in light scattering}      
     in: Waves and Particles in Light and Matter, eds.: A.Garuccio and A.van der Merwe 
     (Plenum, New York, 1994); also e-print gr-qc/9410036. 
\bibitem{DodHal03} P. J. Dodd and J. J. Halliwell, Phys. Rev. {\bf D67}, 105018 (2003). 
\bibitem{Horetal03} K. Hornberger, S. Uttenthaler, B. Brezger, L. Hackerm\"uller, M. Arndt, and A. Zeilinger, Phys. Rev. Lett. {\bf 90}, 160401 (2003).
\bibitem{Dio95} L. Di\'osi, Europhys.Lett. {\bf 30}, 63 (1995). 
\bibitem{Vac01} B. Vacchini, Phys. Rev. {\bf E63}, 066115 (2001)
\bibitem{Hor06} K. Hornberger, Phys. Rev. Lett. {\bf 97}, 060601 (2006).
\bibitem{HorSip03} K. Hornberger and J. E. Sipe, Phys. Rev. {\bf A68}, 012105 (2003).
\bibitem{Adl06} S. L. Adler J. Phys. A: Math. Gen. {\bf 39} 14067 (2006).   

\bibitem{HorVac08} K. Hornberger and B. Vacchini, Phys. Rev. A {\bf 77}, 022112 (2008).
\bibitem{VacHor09} B. Vacchini and K. Hornberger, arXiv:0904.3911v1 [quant-ph] (2009).
\bibitem{KamCre09} I. Kamleitner and J. Cresser, arXiv:0905.3233v1 [quant-ph] (2009).
\bibitem{Dek81} H. Dekker, Phys. Rep. {\bf 80}, 1 (1981).
\bibitem{Jacetal99Beretal06} K. Jacobs, I. Tittonen, H. M. Wiseman and S. Schiller Phys. Rev. {\bf A60}, 538 (1999);
                             J.Z.Bern\'ad, L. Di\'osi and T.Geszti, Phys. Rev. Lett. {\bf 97}, 250404 (2006).
\end{thebibliography}
\end{document}